\documentclass[reprint,superscriptaddress,amsmath,amssymb,aps,prl]{revtex4-1}
\usepackage{graphicx}
\usepackage{dcolumn}
\usepackage{bm}
\usepackage{upgreek}
\usepackage{xcolor}

\begin{document}

\newcommand{\be}{\begin{equation}}
\newcommand{\ee}{\end{equation}}
\newcommand{\bea}{\begin{eqnarray}}
\newcommand{\eea}{\end{eqnarray}}
\newcommand{\HH}{{\cal H}}
\newcommand{\LL}{{\cal L}}
\newcommand{\KK}{{\cal K}}
\newcommand{\GG}{{\sf G}}
\newcommand{\tr}{{\rm tr\/}}
\newcommand{\p}{\partial}
\newcommand{\s}{\sigma}
\newcommand{\la}{\langle}
\newcommand{\ra}{\rangle}
\newcommand{\lb}{\left[}
\newcommand{\rb}{\right]}
\newcommand{\lp}{\left(}
\newcommand{\rp}{\right)}
\newcommand{\Ztwo}{$\mathbb Z_2$ }
\renewcommand{\vec}[1]{{\bf #1}}
\def\nn{\nonumber\\}

\preprint{APS/123-QED}
 
\title{$\mathbb{Z}_2$ phases and Majorana spectroscopy in paired Bose-Hubbard chains}
\author{Smitha Vishveshwara}
\affiliation{Department of Physics, University of Illinois at Urbana-Champaign, Urbana, Illinois 61801-3080, USA}
\email{smivish@illinois.edu}
\author{David M. Weld}
\affiliation{Department of Physics, University of California, Santa Barbara, California 93106, USA}

\begin{abstract} 
We investigate the Bose-Hubbard chain in the presence of nearest-neighbor pairing. The pairing term gives rise to an unusual gapped \Ztwo Ising phase that has number fluctuation but no off-diagonal long range order. This phase has a strongly correlated many-body doubly degenerate ground state which is effectively a gap-protected macroscopic qubit. In the strongly interacting limit, the system can be mapped onto an anisotropic transverse spin chain, which in turn can be mapped to the better-known fermionic sister of the paired Bose-Hubbard chain: the Kitaev chain which hosts zero-energy Majorana bound states. While corresponding phases in the fermionic and bosonic systems have starkly different wavefunctions, they share identical energy spectra. We describe a possible cold-atom realization of the paired Bose-Hubbard model in a biased zig-zag optical lattice with reservoir-induced pairing, opening a possible route towards experimental Kitaev chain spectroscopy. 
\end{abstract}

\maketitle
While $p$-wave fermionic pairing is a subject of intense ongoing investigations, analogous bosonic phenomena are less well explored. Here we pose and address a variety of questions regarding pairing in spinless 1D bosonic systems: What new phases emerge due to pairing? What are the effects of interactions? What parallels exist to 1D fermionic counterparts? The fermionic Kitaev chain hosts gapped topological phases and mid-gap Majorana states; do bosonic systems exhibit related features?

\begin{figure}[t!]
\centering
\includegraphics[width=0.9\columnwidth]{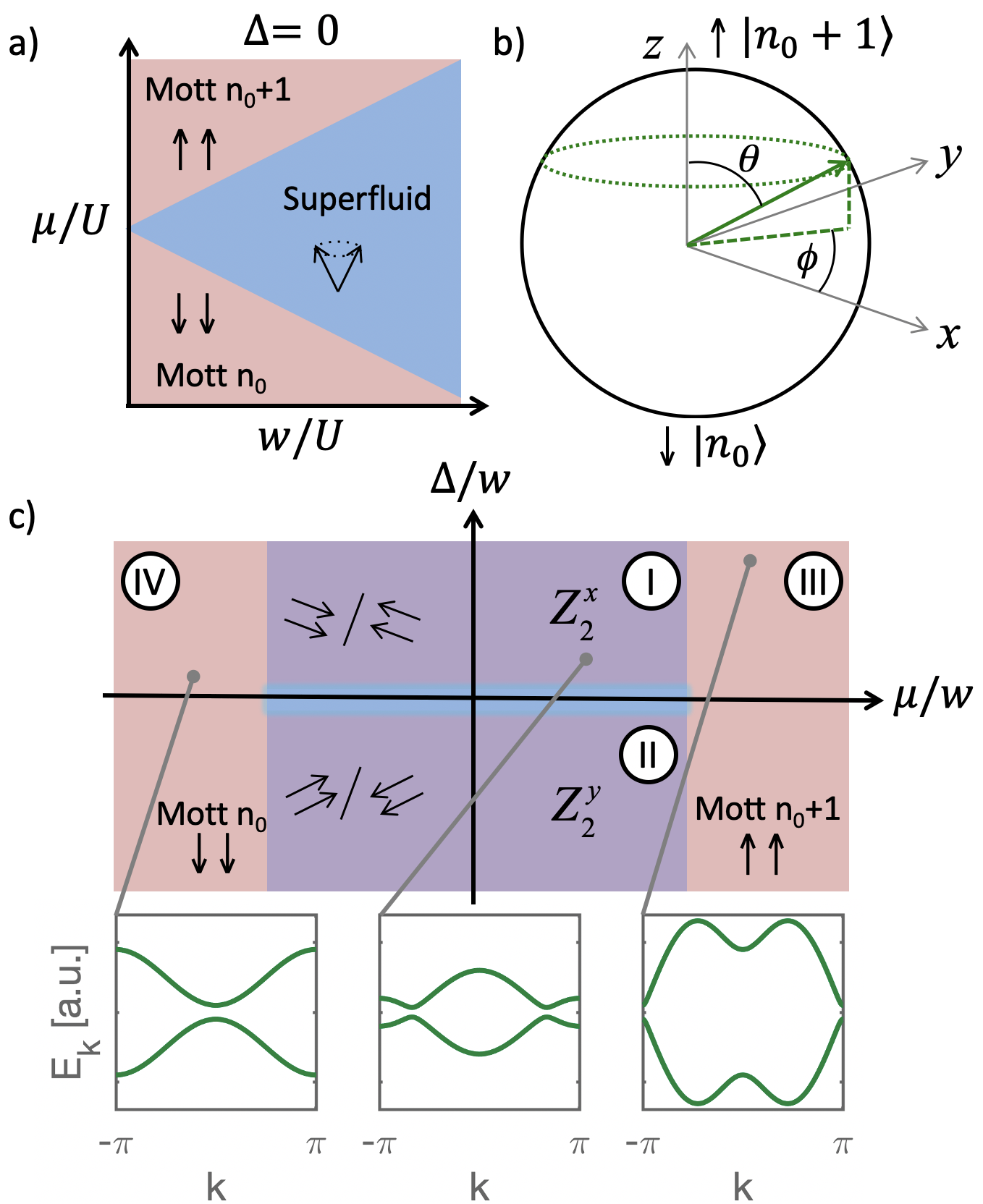}
\caption{Phases of the paired Bose-Hubbard model. {\bf a)}~Superfluid and Mott phases for large $U$ and $\Delta=0$. {\bf b)} The pseudospin Bloch sphere, with Mott phases at the poles and superfluid phases breaking the associated $U(1)$ symmetry. {\bf c)} Finite $\Delta$ creates a new phase that breaks this symmetry down to $Z_2$. Energy spectra, shown for a few representative points, are identical to those of the fermionic Kitaev chain.  }
\label{fig:phases}
\end{figure}

The Bose-Hubbard model offers fertile ground to explore these questions; its rich phase diagram has been characterized for decades~\cite{MPAFbosons,sachdevbook,zoller-hubbard,PhysRevLett.92.050402,Sheshadri_1993}, sustaining a productive interplay between theory and experiment since the advent of direct realizations using ultracold atoms in optical lattices~\cite{zoller-hubbard,greinerSFMI}. The phase diagram hosts a gapless superfluid phase exhibiting number fluctuations and condensation as well as multiple gapped Mott phases. A commonly employed mapping between site occupations and spins serves a fruitful source of physical intuition and a connection to models of magnetism. The description reduces to a spin-$1/2$ system in the limit of strong interactions where the occupation number on each site can only fluctuate between particular numbers $n_0$ and $n_0+1$~\cite{sachdevbook, Sheshadri_1993,PhysRevA.75.063622,10.1143/PTP.16.569,PhysRevB.47.342,PhysRevLett.89.250404,PhysRevB.44.10328}. Here, we focus on a homogeneous 1D chain of bosons  hopping between neighboring lattice sites with tunneling strength $w$ and an on-site interaction $U$. The  chemical potential $\mu$ is determined by the total number of bosons. The additional crucial ingredient is a pairing contribution that creates and annihilates pairs of bosons on neighboring sites with strength $\Delta$. This chain is described by the Hamiltonian

\bea\label{eq:pairedBH}
\HH & = & \HH_{BH}+\HH_{\Delta} \\
\HH_{BH} & = & -w \sum_{\la ij\ra}(b^\dag_ib_j +  h.c.) + \sum_i(\frac{U}{2}\hat{n}_i(\hat{n}_i-1)-\mu \hat{n_i}), \nonumber \\
\HH_{\Delta} & = & -\Delta \sum_{\la ij\ra}( b^\dag_ib^\dag_j + h.c.). \nonumber 
\eea
Here $\la ij\ra$ denotes a sum over nearest neighbor sites, $b_i^\dag$ ($b_i$) is the creation (annihilation) operator for a boson on site $i$, and $\hat{n_i}=b_i^\dag b_i$ is the number operator. The  effect of the pairing term, $\Delta$, on 1D Bose-Hubbard physics is the main subject of this work.

Combining established insights stemming from boson-to-spin and spin-to-fermion maps in one-dimension, we demonstrate that pairing gives rise to remarkable new features in strongly interacting Bose-Hubbard chains. Most prominently, as depicted in Fig.~\ref{fig:phases}, we find that pairing reduces the $U(1)$ symmetry of the unpaired system to an Ising symmetry, leading to a unique \Ztwo symmetry-broken phase. As with the usual condensate, which corresponds to the $U(1)$ symmetry-broken phase, this phase exhibits number fluctuations on each site. However, it is gapped and lacks the Goldstone modes associated with an ordinary superfluid. Moreover, its ground state is doubly degenerate, reflecting two very different superpositions of bosonic occupation on every site. Effectively, this subspace forms a gap-protected macroscopic qubit. We note a close relation to non-Abelian Majorana bound state physics: while this \Ztwo phase is not a topological phase that hosts such states, the two phases can be connected via a specific non-local mapping. Furthermore, since the energy spectrum of the Hamiltonian $\HH$ is identical to that of the fermionic Kitaev chain, a Bose-Hubbard realization would enable experimental measurement of the spectrum for the entire phase diagram. In what follows, we present our analyses of these features, beginning for concreteness by briefly describing a possible  realization. 

Experimental access to the paired Bose-Hubbard model is most naturally attained using controlled systems of photons or bosonic atoms. Such systems have attracted theoretical interest due to the close relationship to the Kitaev chain: proposals and investigations include employing photonic modes as analogues to Majorana states~\cite{photonicmajorana1,photonicmajorana2,photonicmajorana3}, and ultracold fermions~\cite{topocoldatomsreview,Nascimb_ne_2013,majoranacoldatomwires,PhysRevLett.111.173004} and bosons~\cite{eckardtanyons} for the realization of anyonic modes. In the context of cold atoms, experimental access to the bosonic Kitaev chain analogue is more straightforward than in the original fermionic case: an optical lattice configured as a biased zigzag chain offers a possible natural experimental realization of the Hamiltonian $\HH$ in Eq.~\ref{eq:pairedBH}. Fig.~2 diagrams a section of such a potential, which can be created in a four-beam monochromatic optical lattice with a controllable relative phase between the beams. An appropriate choice of phases yields an array of quasi-1D zig-zag ladders having a tunable chemical potential offset $\mu_0$ between the legs. We define the higher-energy leg as our system of interest and the lower-energy leg as the ``reservoir chain.'' In a tight-binding picture, the hopping term $w$ is achieved through tunneling along the upper chain.  The pairing term $\Delta$ is realized by setting $\mu_0=U_r$ where $U_r$ is the interaction energy of two bosons on the reservoir chain. At this offset, single-particle tunneling between the legs of the ladder is non-resonant, and resonant {\it pair} tunneling dominates. Similar resonant or near-resonant chemical potential offsets have been shown to enable control of superexchange dynamics and realization of spin Hamiltonians in optical lattices~\cite{sachdev-tiltedlatt1,greiner-AFspinchains,nagerl-tiltedlatt,ketterle-tiltedlatt}. To maintain constant pairing $\Delta$ and an associated phase which we arbitrarily set to 0, the tunneling rate within the reservoir chain should be set at a value sufficient to stabilize superfluidity. Asymmetrical interchain hopping rates can be tuned by adjusting the lattice phase.

\begin{figure}[t]
\centering
\includegraphics[width=0.8\columnwidth]{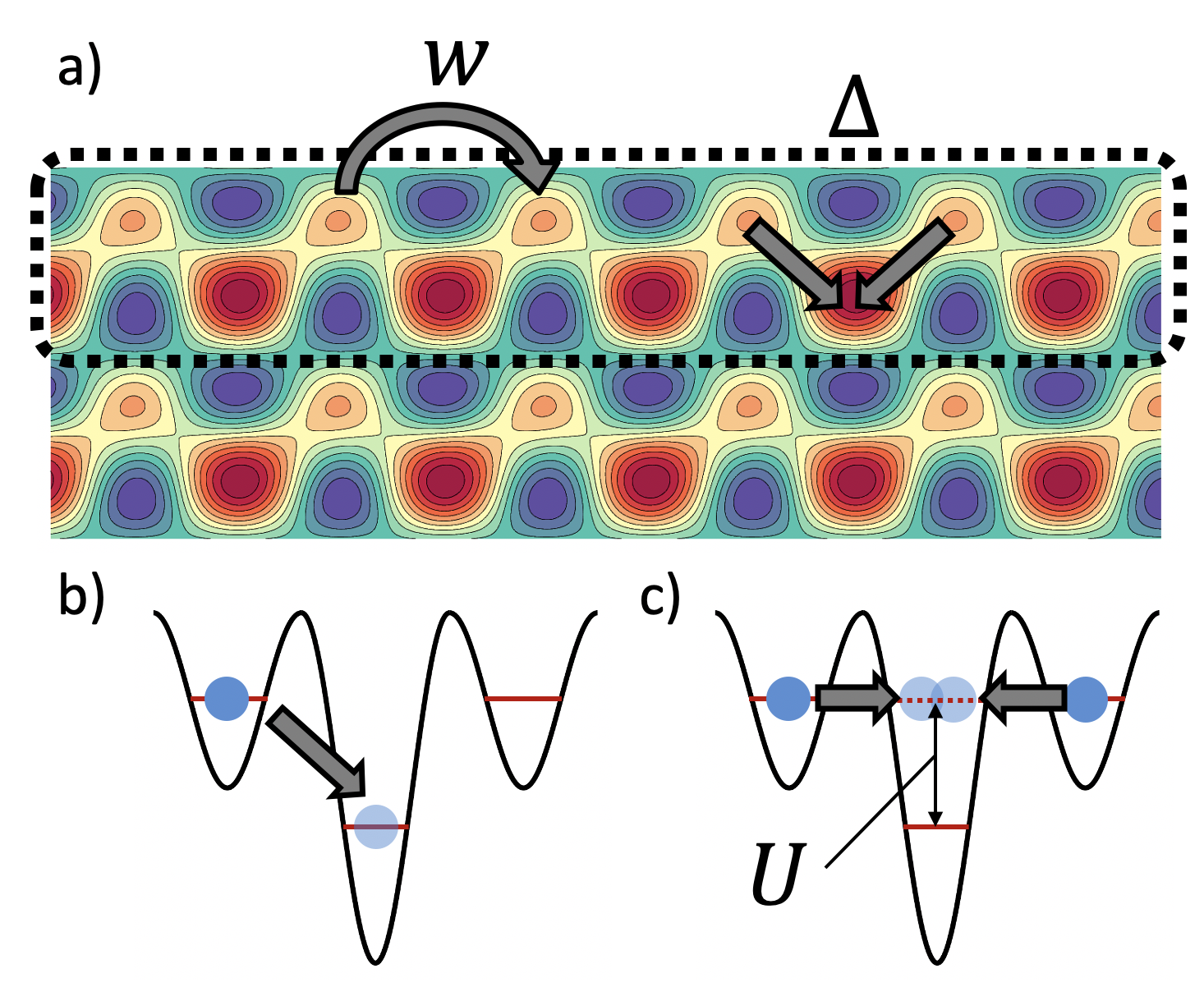}
\caption{Potential optical lattice realization of the paired 1D BH model.  \textbf{a:} Array of zigzag ladder potentials created by four-beam optical lattice with tunable interbeam phase. Hopping and pairing terms in a single ladder (boxed) are schematically indicated.  Here the lower leg of each ladder is the ``reservoir chain.''  \textbf{b:} Schematic of off-resonant single-particle interchain hopping.  \textbf{c:} Schematic of resonant pair hopping.}
\label{lattfig}
\end{figure}

Now we show that approaches commonly employed to analyze the standard Bose-Hubbard chain can provide insight into the effects of the additional pairing term. We focus on the regime of $U$ and $\mu$ such that  the average density on each site lies between $n_0$ and $n_0+1$. For larger enough interaction strength, $w/U\ll 1$,  the Hilbert space can  be restricted to the number-basis states $|n_0\ra$ and $|n_0+1\ra$ at each site. The most energetically relevant excluded states, $|n_0-1\ra$ and $|n_0+2\ra$, would contribute corrections of order of $w^2/U$. Following truncation, the bosonic chain can be mapped to an $XY$ (pseudo-)spin-$1/2$ chain in a transverse field \cite{PhysRevA.75.063622,sachdev-tiltedlatt1,PhysRevB.44.10328,greiner-AFspinchains,sachdevbook}. 
The truncated Hilbert space may be represented by the spin-$1/2$
states, $|n_0+1\ra=|\uparrow\ra$ and
$|n_0\ra=|\downarrow\ra$, the eigenstates of the operator $s^z$ having eigenvalues
$\pm 1/2$. The tunneling term in the Bose-Hubbard Hamiltonian can be identified
with raising and lowering spin-$1/2$ operators, $s^+$ and $s^-$, such that $
b^\dagger_i b_j\to (n_0+1)s_i^+ s_j^-$. The interaction and the potential energy terms
are diagonal in the number basis at each site, and the boson number operator, $\hat{n}_i$,  can be expressed in terms of the spin-1/2 matrix
$s^z$: $\hat n \to n_0+1/2+s^z$. In the truncated spin-$1/2$ Hilbert space, the one-dimensional version of the paired  Bose-Hubbard  Hamiltonian of Eq.~\ref{eq:pairedBH} takes the form 
\be\label{eq:Ham_spin}
\HH_S=-\sum_{\la ij\ra}\lp J_x s^x_i s^x_j+J_y s^y_i s^y_j\rp - \sum_i
h s^z_i, 
\ee
where we have the identifications $J_x\leftrightarrow 2(n_0+1)(w + \Delta)$, $J_y\leftrightarrow 2(n_0+1)(w-\Delta)$, and $h \leftrightarrow \mu-Un_0$. As shown in Fig.~\ref{fig:phases}a and \ref{fig:phases}b, in the standard case without pairing, Mott states having integer boson filling $n_0$ and $n_0+1$ correspond to gapped phases in which spins are polarized along the $\pm\hat{z}$ directions, while the gapless superfluid phase at intermediate fillings corresponds to a ferromagnetic state which spontaneously breaks $U(1)$ symmetry.

The presence of the pairing term renders the $XY$ couplings anisotropic, indicative of the aforementioned reduction of the gapless $U(1)$ phase  to a gapped phase of the Ising \Ztwo universality class. In order to study this effect of the pairing term, it is useful to recall features of the standard unpaired case,  where $\Delta=0$ and $J_x=J_y\equiv J$. The $U(1)$ symmetry associated with particle number can  be seen in  Eq. \ref{eq:pairedBH} by noting its invariance under the transformation $b^{\dagger}\rightarrow b^{\dagger} e^{i\phi},b\rightarrow b e^{-i\phi}$.  The superfluid phase associated with the breaking of this symmetry exists even beyond the spin-1/2 limit where $w/U\ll 1$; for large enough tunneling, it includes occupation numbers beyond the $n_0$ and $n_0+1$ Mott lobes. In the spin-1/2 limit described by Eq. \ref{eq:Ham_spin}, a large enough transverse field strength $h$,  or equivalently,  deviation of the chemical potential $\mu$ from the value $Un_0$, tips the system into the $n_0$ or $n_0+1$ Mott states, depending on the sign of the field. For field strengths lower than this critical value, the system is in the superfluid state. 
At a mean-field level, which is more appropriate for higher dimensions but provides good insights into the phases for any dimension, this behavior can be captured by replacing the pseudospin on each site with an expectation value parametrized by $\langle \vec{S}_i \rangle = \frac{1}{2} (\sin\theta\cos\phi, \sin\theta\sin\phi, \cos\theta )$. The Hamiltonian of Eq.~\ref{eq:Ham_spin} then takes the form of a spin on each site in an effective magnetic field, $H_{\mathrm{mf}}=-\vec{S}_i\cdot\vec{B}_{\mathrm{mf}}$, where for a chain, $\vec{B}_{\mathrm{mf}}=(J \sin\theta\cos\phi, J \sin\theta\sin\phi, h)$. The equilibrium configuration of the pseudospin corresponds to minimizing the associated energy, yielding $\cos\theta=h/J$ and no constraints on the value of $\phi$.  In the Mott phase, the pseudospins are completely polarized along the $z$ direction, i.e. $\la s^z_i\ra=\pm 1/2$,
allowing the identification of
 $\mu_{\pm}=Un_0\pm 2 w(n_0+1)$, the values of the chemical potential
at the boundaries of the Mott states having $n_0$ and $n_0+1$ bosons per site. In the superfluid phase, the mean-field ground state comprises a superposition on each site: 
\be\label{eq:psi}
|\psi\ra_i=\cos\theta|n_0\ra_i+e^{i\varphi}\sin\theta|n_0+1\ra_i.
\ee
 Spontaneous $U(1)$ symmetry breaking entails making a choice of the continuous parameter $\varphi$, giving rise to a gapless Goldstone mode in the excitation spectrum. 

Compared with this unpaired Bose-Hubbard chain, in which number fluctuations on a given site are due to nearest neighbor hopping, the pairing term in Eq.~\ref{eq:pairedBH} allows for the insertion and depletion of bosons in pairs at neighboring sites. The system no longer respects the invariance under $b^{\dagger}\rightarrow b^{\dagger} e^{i\phi},b\rightarrow b e^{-i\phi}$  associated with number conservation. The only exception is for the choices $\phi=0,\pi$, reflecting pariwise processes as opposed to those of single bosons. Thus, in the presence of the pairing term,  $U(1)$ symmetry is reduced to \Ztwo Ising symmetry. In the spin-chain limit of Eq. \ref{eq:Ham_spin}, the reduction to the Ising symmetry is reflected in the identification $\Delta \leftrightarrow J_x-J_y$.  The inequality of $J_x$ and $J_y$ leads to the reduction to Ising symmetry; for $J_x>J_y$, the ferromagnetic coupling favors ordering along the eigenstates of the Pauli spin $\s^x$ on each site and for $J_x<J_y$ along $\s^y$. In terms of spin symmetry properties, one can see that Eq.~\ref{eq:Ham_spin} is invariant under the transformation $s_i^x\rightarrow-s_i^x; s_i^y\rightarrow-s^i_y;s_i^z\rightarrow s_i^z$. 

Repeating the mean-field argument above in the presence of pairing establishes salient features of the \Ztwo phase. Energy minimization is no longer independent of $\phi$ due to the presence of the non-vanishing $\Delta$ term. For $J_x>J_y$, the phase is pinned to values $\phi=0,\pi$, while for $J_x<J_y$, it is pinned to $\phi=\pi/2,3\pi/2$. In these two cases, the average density, captured by the expectation value $\langle s^z_i \rangle$, now varies as $\cos\theta=h/J_x$ and $\cos\theta=h/J_y$, respectively.  Hence, once more, the system enters into the regular  $n_0\: (\theta=\pi/2)$ or $n_0+1\: (\theta=0)$ Mott phases for large enough effective field. However, for couplings that allow a solution in the range $0<\theta<\pi/2$ , the symmetry-broken \Ztwo phase is different from the usual condensate phase.  On the one hand, it too exhibits number fluctuations such as those depicted in Eq.~\ref{eq:psi}. On the other, the pairing term pins the phase such that no continuous symmetry is broken. As seen from the mean-field wavefunction, depending on the relative magnitudes of the anisotropic couplings,  the phase $\phi$  is pinned to one of two possible values that are consistent with eigenstates of $s^x$ or $s^y$, respectively, translating to two macroscopically different ground state wavefunctions for the bosonic chain. Furthermore, the absence of $U(1)$ symmetry breaking implies the absence of a low-lying Goldstone mode; the system is gapped. 

Thus, a remarkable feature of this bosonic \Ztwo phase is that its correlated many-body ground state is doubly degenerate and protected by a gap. As a simple example, in the limit of equal hopping and pairing, $\Delta=w$, the system is exactly in the transverse Ising limit $J_y=0$. If the system is now set to half filling, $h=\mu-Un_0=0$, then the mean-field states described above become the exact ground states with all spins either pointing along $+x$ or $-x$. In terms of bosons, every site has equal superposition of $n_0$ and $n_0+1$ bosons and the degeneracy corresponds to all sites having a symmetric or antisymmetric superposition. 

Having discussed the possible \Ztwo phases of the paired Bose-Hubbard chain and their ground state properties, we proceed to describe the entire  energy spectrum in the regime of interest. To this end, the transverse $XY$ spin chain of Eq.~\ref{eq:Ham_spin} can be diagonalized via the standard Jordan-Wigner transformation~\cite{LIEB1961407,sachdevbook,PhysRevB.88.165111}. Here, the spin operators on a given site $j$ along the chain are expressed in terms of  fermionic creation and annihilation operators ($f^{\dagger},f$) through the transformation $s^x_j = f_j^{\dagger}f_j-1/2$, $s^+_j = f^{\dagger}_j e^{-i\pi\sum_{l<j} N_l}$, where $N_l$ is the fermion occupation number on site $l$. The spin chain Hamiltonian, in terms of the fermions, transforms to a one-dimensional tight-binding representation of a p-wave superconductor, known as the Kitaev chain~\cite{kitaev1,Zhang:2019bl,PhysRevX.6.031016,Sau:2020aj}, whose chief feature is that it supports topologically robust isolated Majorana fermionic bound states. The Kitaev chain consists of spinless fermions experiencing nearest neighbor hopping $\tilde{w}$ and pairing $\tilde{\Delta}$, and is described by the Hamiltonian
\begin{align}
\label{eq:kitaevchain}
\HH_F &= \sum_{j=1}^{N-1} \Big[ -\tilde{w}\left(f_j^\dagger f_{j+1}+f_{j+1}^\dagger f_j\right) \\
 &+\tilde{\Delta}\left(f_jf_{j+1}+f_{j+1}^\dagger f_j^\dagger \right)\Big] - \tilde{\mu}\sum_{j=1}^N\left(f_j^\dagger f_j - \frac{1}{2}\right). \nonumber
\end{align}
In terms of the original Bose-Hubbard parameters, we can identify $\tilde{w}=w (n_0+1)$ and $\tilde{\Delta}=-\Delta(n_0+1)$. The effective magnetic field experienced by the transverse spin chain translates to $\tilde{\mu}=\mu-Un_0$, namely, an effective chemical potential for fermions.

As can be seen by comparing the fermionic Hamiltonian above with the paired bosonic Hamiltonian of Eq.\ref{eq:pairedBH}, nearest neighbor hopping and pairing terms in one system directly translate to those in the other. While the two systems are very similar in form, the nature of the phases and states they exhibit are radically different. The \Ztwo aspect, while common to both, reflects local order in the bosonic system and topological order in the fermionic system. The bosonic \Ztwo phase discussed above corresponds to a topological phase in the fermionic chain that supports Majorana bound states at its ends. The ground state degeneracy is common to both systems, and in the fermionic case, corresponds to the pair of Majorana end bound states forming a Dirac fermionic state that can either be occupied or unoccupied. The \Ztwo degree of freedom is thus associated with electron parity. Spontaneous symmetry breaking in the topological phase thus corresponds to picking odd or even parity, or a specific superposition. The Mott-insulating phase in the bosonic system translates  in the fermionic system to a topologically trivial non-degenerate phase having no isolated Majorana bound states.

Not only is the ground state double degeneracy common to the symmetry-broken \Ztwo phases in the bosonic and fermionic systems described by Eq.~\ref{eq:pairedBH} and Eq.~\ref{eq:kitaevchain}, respectively, as expected of basis-independent properties, the full energy spectrum is identical across the entire range of parameters. Specifically, diagonalizing the quadratic fermionic Hamiltonian by transforming into momentum space gives the energy dispersion
\be
E_k ~=~ \pm ~\sqrt{(2\tilde{w} \cos k + \tilde{\mu})^2 ~+~ 4 \tilde{\Delta}^2  \sin^2 k}.
\label{eq:dispersion} 
\ee
The dispersion is gapped across the range of parameters, except for the lines i) $\tilde{\mu}= 2\tilde{w}$, ii) $\tilde{\mu}= -2\tilde{w}$ and iii) $\tilde{\Delta}=0, |\tilde{\mu}|<2\tilde{w}$. Along these lines, the dispersion vanishes at specific momenta; the lines demarcate phase boundaries between different gapped phases. In terms of paired boson physics, the line along $\tilde{\Delta}=0$ corresponds to the standard superfluid-Mott insulator phase diagram in the limit of large interaction for a given hopping strength $w$. As depicted in Fig.\ref{fig:phases},  Phases I and II correspond to the doubly-degenerate \Ztwo phases.  Phases III and IV correspond to the bosonic Mott phases and their  topologically trivial fermionic counterparts. For any given point in the phase diagram, Eq.~\ref{eq:dispersion} provides the energy dispersion, which can be easily modified to a discrete spectrum for finite-sized systems. The momentum modes, while corresponding to free fermions, are highly non-trivial in terms of bosons due to the non-local mapping between the two bases. The energy spectrum, however, is common to both. Furthermore, in the \Ztwo phases, the two degenerate states populate the center of the gap. Finite size effects are expected to split this degeneracy.

Returning to experimental realizations, we note that all major ingredients of the proposed scheme have already been demonstrated in cold atom experiments: tunable hopping, interaction-induced resonant tunneling, and of course tight-binding Hamiltonians~\cite{greinerSFMI,greiner-AFspinchains}. The standard Bose-Hubbard model has been a particularly fruitful focus for cold atom experimental work: since the initial realization of the superfluid-Mott insulator phase transition with ultracold rubidium in an optical lattice~\cite{greinerSFMI}, a wide variety of experiments have probed aspects ranging from compressibility~\cite{chinmottplateaux} to string ordering~\cite{bloch-stringorder}. A cold-atom manifestation of the paired Bose-Hubbard model would build upon these and related developments of advanced characterization techniques such as spectroscopic probes~\cite{clockshiftmott,esslinger_SMFI_latticemodulation,KetterleMottspectroscopy}.  For revealing the Kitaev chain spectrum, modulation spectroscopy of atoms in optical lattices would be a natural approach.  Bragg spectroscopy can reveal energy-quasimomentum relations~\cite{inguscio-bragg,sengstock-bragg}, complemented with new techniques based on position-space Bloch oscillations~\cite{RSBO-PRL,warpdrivePRL} and exotic drives~\cite{phasonPRL}. 
In principle, wavefunction features of the \Ztwo phase highlighted in this work  could be observed in interferometric signatures in time-of-flight images of a multiple-chain sample.

In summary, we have investigated the physics of  Bose-Hubbard chains with nearest-neighbor pairing, motivated by a possible experimental realization using an interaction anti-blockade in zig-zag optical lattices. Physical insights into the resulting features are revealed by mapping to an anisotropic XY spin chain and from there to a fermionic chain. The model exhibits unusual gapped \Ztwo phases and a spectrum identical to that of the Kitaev chain, including doubly-degenerate zero-energy Majorana bound states.  Future work suggested by these results could include considerations of larger spin, higher dimensions, and inhomogeneities, as well as the exploration of possible connections to $XY$ lattice models of fractionalization in gapless $U(1)$ versus gapped $\mathbb{Z}_N$ phases~\cite{seibergarxiv}.

\begin{acknowledgements}
We are grateful for illuminating discussions with Jason Alicea, Assa Auerbach, Wade DeGottardi, Matthew Fisher, Anthony Leggett, Diptiman Sen, and Norman Yao. This research was supported in part by the National Science Foundation under Grant No. NSF PHY-1748958. DW acknowledges support during the conceptual stages of this work from the Army Research Office (MURI W911NF1710323). SV acknowledges UC San Diego's gracious hospitality through the Margaret Burbidge Visiting Professorship Award during the inital parts of this work. This material is based upon work supported by the U.S. Department of Energy, Office of Science, National Quantum Information Science Research Centers.  
\end{acknowledgements}


\begin{thebibliography}{40}%
\makeatletter
\providecommand \@ifxundefined [1]{%
 \@ifx{#1\undefined}
}%
\providecommand \@ifnum [1]{%
 \ifnum #1\expandafter \@firstoftwo
 \else \expandafter \@secondoftwo
 \fi
}%
\providecommand \@ifx [1]{%
 \ifx #1\expandafter \@firstoftwo
 \else \expandafter \@secondoftwo
 \fi
}%
\providecommand \natexlab [1]{#1}%
\providecommand \enquote  [1]{``#1''}%
\providecommand \bibnamefont  [1]{#1}%
\providecommand \bibfnamefont [1]{#1}%
\providecommand \citenamefont [1]{#1}%
\providecommand \href@noop [0]{\@secondoftwo}%
\providecommand \href [0]{\begingroup \@sanitize@url \@href}%
\providecommand \@href[1]{\@@startlink{#1}\@@href}%
\providecommand \@@href[1]{\endgroup#1\@@endlink}%
\providecommand \@sanitize@url [0]{\catcode `\\12\catcode `\$12\catcode
  `\&12\catcode `\#12\catcode `\^12\catcode `\_12\catcode `\%12\relax}%
\providecommand \@@startlink[1]{}%
\providecommand \@@endlink[0]{}%
\providecommand \url  [0]{\begingroup\@sanitize@url \@url }%
\providecommand \@url [1]{\endgroup\@href {#1}{\urlprefix }}%
\providecommand \urlprefix  [0]{URL }%
\providecommand \Eprint [0]{\href }%
\providecommand \doibase [0]{http://dx.doi.org/}%
\providecommand \selectlanguage [0]{\@gobble}%
\providecommand \bibinfo  [0]{\@secondoftwo}%
\providecommand \bibfield  [0]{\@secondoftwo}%
\providecommand \translation [1]{[#1]}%
\providecommand \BibitemOpen [0]{}%
\providecommand \bibitemStop [0]{}%
\providecommand \bibitemNoStop [0]{.\EOS\space}%
\providecommand \EOS [0]{\spacefactor3000\relax}%
\providecommand \BibitemShut  [1]{\csname bibitem#1\endcsname}%
\let\auto@bib@innerbib\@empty
\bibitem [{\citenamefont {{Fisher}}\ \emph {et~al.}(1989)\citenamefont
  {{Fisher}}, \citenamefont {{Weichman}}, \citenamefont {{Grinstein}},\ and\
  \citenamefont {{Fisher}}}]{MPAFbosons}%
  \BibitemOpen
  \bibfield  {author} {\bibinfo {author} {\bibfnamefont {M.~P.~A.}\
  \bibnamefont {{Fisher}}}, \bibinfo {author} {\bibfnamefont {P.~B.}\
  \bibnamefont {{Weichman}}}, \bibinfo {author} {\bibfnamefont
  {G.}~\bibnamefont {{Grinstein}}}, \ and\ \bibinfo {author} {\bibfnamefont
  {D.~S.}\ \bibnamefont {{Fisher}}},\ }\href {\doibase 10.1103/PhysRevB.40.546}
  {\bibfield  {journal} {\bibinfo  {journal} {\prb}\ }\textbf {\bibinfo
  {volume} {40}},\ \bibinfo {pages} {546} (\bibinfo {year} {1989})}\BibitemShut
  {NoStop}%
\bibitem [{\citenamefont {Sachdev}(2011)}]{sachdevbook}%
  \BibitemOpen
  \bibfield  {author} {\bibinfo {author} {\bibfnamefont {S.}~\bibnamefont
  {Sachdev}},\ }\href@noop {} {\emph {\bibinfo {title} {Quantum Phase
  Transitions}}}\ (\bibinfo  {publisher} {Cambridge},\ \bibinfo {year}
  {2011})\BibitemShut {NoStop}%
\bibitem [{\citenamefont {Jaksch}\ \emph {et~al.}(1998)\citenamefont {Jaksch},
  \citenamefont {Bruder}, \citenamefont {Cirac}, \citenamefont {Gardiner},\
  and\ \citenamefont {Zoller}}]{zoller-hubbard}%
  \BibitemOpen
  \bibfield  {author} {\bibinfo {author} {\bibfnamefont {D.}~\bibnamefont
  {Jaksch}}, \bibinfo {author} {\bibfnamefont {C.}~\bibnamefont {Bruder}},
  \bibinfo {author} {\bibfnamefont {J.~I.}\ \bibnamefont {Cirac}}, \bibinfo
  {author} {\bibfnamefont {C.~W.}\ \bibnamefont {Gardiner}}, \ and\ \bibinfo
  {author} {\bibfnamefont {P.}~\bibnamefont {Zoller}},\ }\href@noop {}
  {\bibfield  {journal} {\bibinfo  {journal} {Phys. Rev. Lett.}\ }\textbf
  {\bibinfo {volume} {81}},\ \bibinfo {pages} {3108} (\bibinfo {year}
  {1998})}\BibitemShut {NoStop}%
\bibitem [{\citenamefont {Kuklov}\ \emph {et~al.}(2004)\citenamefont {Kuklov},
  \citenamefont {Prokof'ev},\ and\ \citenamefont
  {Svistunov}}]{PhysRevLett.92.050402}%
  \BibitemOpen
  \bibfield  {author} {\bibinfo {author} {\bibfnamefont {A.}~\bibnamefont
  {Kuklov}}, \bibinfo {author} {\bibfnamefont {N.}~\bibnamefont {Prokof'ev}}, \
  and\ \bibinfo {author} {\bibfnamefont {B.}~\bibnamefont {Svistunov}},\ }\href
  {\doibase 10.1103/PhysRevLett.92.050402} {\bibfield  {journal} {\bibinfo
  {journal} {Phys. Rev. Lett.}\ }\textbf {\bibinfo {volume} {92}},\ \bibinfo
  {pages} {050402} (\bibinfo {year} {2004})}\BibitemShut {NoStop}%
\bibitem [{\citenamefont {Sheshadri}\ \emph {et~al.}(1993)\citenamefont
  {Sheshadri}, \citenamefont {Krishnamurthy}, \citenamefont {Pandit},\ and\
  \citenamefont {Ramakrishnan}}]{Sheshadri_1993}%
  \BibitemOpen
  \bibfield  {author} {\bibinfo {author} {\bibfnamefont {K.}~\bibnamefont
  {Sheshadri}}, \bibinfo {author} {\bibfnamefont {H.~R.}\ \bibnamefont
  {Krishnamurthy}}, \bibinfo {author} {\bibfnamefont {R.}~\bibnamefont
  {Pandit}}, \ and\ \bibinfo {author} {\bibfnamefont {T.~V.}\ \bibnamefont
  {Ramakrishnan}},\ }\href {\doibase 10.1209/0295-5075/22/4/004} {\bibfield
  {journal} {\bibinfo  {journal} {Europhysics Letters ({EPL})}\ }\textbf
  {\bibinfo {volume} {22}},\ \bibinfo {pages} {257} (\bibinfo {year}
  {1993})}\BibitemShut {NoStop}%
\bibitem [{\citenamefont {Greiner}\ \emph {et~al.}(2002)\citenamefont
  {Greiner}, \citenamefont {Mandel}, \citenamefont {Esslinger}, \citenamefont
  {H{\"a}nsch},\ and\ \citenamefont {Bloch}}]{greinerSFMI}%
  \BibitemOpen
  \bibfield  {author} {\bibinfo {author} {\bibfnamefont {M.}~\bibnamefont
  {Greiner}}, \bibinfo {author} {\bibfnamefont {O.}~\bibnamefont {Mandel}},
  \bibinfo {author} {\bibfnamefont {T.}~\bibnamefont {Esslinger}}, \bibinfo
  {author} {\bibfnamefont {T.~W.}\ \bibnamefont {H{\"a}nsch}}, \ and\ \bibinfo
  {author} {\bibfnamefont {I.}~\bibnamefont {Bloch}},\ }\href@noop {}
  {\bibfield  {journal} {\bibinfo  {journal} {Nature}\ }\textbf {\bibinfo
  {volume} {415}},\ \bibinfo {pages} {39} (\bibinfo {year} {2002})}\BibitemShut
  {NoStop}%
\bibitem [{\citenamefont {Barankov}\ \emph {et~al.}(2007)\citenamefont
  {Barankov}, \citenamefont {Lannert},\ and\ \citenamefont
  {Vishveshwara}}]{PhysRevA.75.063622}%
  \BibitemOpen
  \bibfield  {author} {\bibinfo {author} {\bibfnamefont {R.~A.}\ \bibnamefont
  {Barankov}}, \bibinfo {author} {\bibfnamefont {C.}~\bibnamefont {Lannert}}, \
  and\ \bibinfo {author} {\bibfnamefont {S.}~\bibnamefont {Vishveshwara}},\
  }\href {\doibase 10.1103/PhysRevA.75.063622} {\bibfield  {journal} {\bibinfo
  {journal} {Phys. Rev. A}\ }\textbf {\bibinfo {volume} {75}},\ \bibinfo
  {pages} {063622} (\bibinfo {year} {2007})}\BibitemShut {NoStop}%
\bibitem [{\citenamefont {Matsubara}\ and\ \citenamefont
  {Matsuda}(1956)}]{10.1143/PTP.16.569}%
  \BibitemOpen
  \bibfield  {author} {\bibinfo {author} {\bibfnamefont {T.}~\bibnamefont
  {Matsubara}}\ and\ \bibinfo {author} {\bibfnamefont {H.}~\bibnamefont
  {Matsuda}},\ }\href {\doibase 10.1143/PTP.16.569} {\bibfield  {journal}
  {\bibinfo  {journal} {Progress of Theoretical Physics}\ }\textbf {\bibinfo
  {volume} {16}},\ \bibinfo {pages} {569} (\bibinfo {year} {1956})}\BibitemShut
  {NoStop}%
\bibitem [{\citenamefont {Bruder}\ \emph {et~al.}(1993)\citenamefont {Bruder},
  \citenamefont {Fazio},\ and\ \citenamefont {Sch\"on}}]{PhysRevB.47.342}%
  \BibitemOpen
  \bibfield  {author} {\bibinfo {author} {\bibfnamefont {C.}~\bibnamefont
  {Bruder}}, \bibinfo {author} {\bibfnamefont {R.}~\bibnamefont {Fazio}}, \
  and\ \bibinfo {author} {\bibfnamefont {G.}~\bibnamefont {Sch\"on}},\ }\href
  {\doibase 10.1103/PhysRevB.47.342} {\bibfield  {journal} {\bibinfo  {journal}
  {Phys. Rev. B}\ }\textbf {\bibinfo {volume} {47}},\ \bibinfo {pages} {342}
  (\bibinfo {year} {1993})}\BibitemShut {NoStop}%
\bibitem [{\citenamefont {Altman}\ and\ \citenamefont
  {Auerbach}(2002)}]{PhysRevLett.89.250404}%
  \BibitemOpen
  \bibfield  {author} {\bibinfo {author} {\bibfnamefont {E.}~\bibnamefont
  {Altman}}\ and\ \bibinfo {author} {\bibfnamefont {A.}~\bibnamefont
  {Auerbach}},\ }\href {\doibase 10.1103/PhysRevLett.89.250404} {\bibfield
  {journal} {\bibinfo  {journal} {Phys. Rev. Lett.}\ }\textbf {\bibinfo
  {volume} {89}},\ \bibinfo {pages} {250404} (\bibinfo {year}
  {2002})}\BibitemShut {NoStop}%
\bibitem [{\citenamefont {Rokhsar}\ and\ \citenamefont
  {Kotliar}(1991)}]{PhysRevB.44.10328}%
  \BibitemOpen
  \bibfield  {author} {\bibinfo {author} {\bibfnamefont {D.~S.}\ \bibnamefont
  {Rokhsar}}\ and\ \bibinfo {author} {\bibfnamefont {B.~G.}\ \bibnamefont
  {Kotliar}},\ }\href {\doibase 10.1103/PhysRevB.44.10328} {\bibfield
  {journal} {\bibinfo  {journal} {Phys. Rev. B}\ }\textbf {\bibinfo {volume}
  {44}},\ \bibinfo {pages} {10328} (\bibinfo {year} {1991})}\BibitemShut
  {NoStop}%
\bibitem [{\citenamefont {Bardyn}\ and\ \citenamefont {\ifmmode \dot{I}\else
  \.{I}\fi{}mamo\ifmmode~\check{g}\else
  \v{g}\fi{}lu}(2012)}]{photonicmajorana1}%
  \BibitemOpen
  \bibfield  {author} {\bibinfo {author} {\bibfnamefont {C.-E.}\ \bibnamefont
  {Bardyn}}\ and\ \bibinfo {author} {\bibfnamefont {A.}~\bibnamefont {\ifmmode
  \dot{I}\else \.{I}\fi{}mamo\ifmmode~\check{g}\else \v{g}\fi{}lu}},\ }\href
  {\doibase 10.1103/PhysRevLett.109.253606} {\bibfield  {journal} {\bibinfo
  {journal} {Phys. Rev. Lett.}\ }\textbf {\bibinfo {volume} {109}},\ \bibinfo
  {pages} {253606} (\bibinfo {year} {2012})}\BibitemShut {NoStop}%
\bibitem [{\citenamefont {McDonald}\ \emph {et~al.}(2018)\citenamefont
  {McDonald}, \citenamefont {Pereg-Barnea},\ and\ \citenamefont
  {Clerk}}]{photonicmajorana2}%
  \BibitemOpen
  \bibfield  {author} {\bibinfo {author} {\bibfnamefont {A.}~\bibnamefont
  {McDonald}}, \bibinfo {author} {\bibfnamefont {T.}~\bibnamefont
  {Pereg-Barnea}}, \ and\ \bibinfo {author} {\bibfnamefont {A.~A.}\
  \bibnamefont {Clerk}},\ }\href {\doibase 10.1103/PhysRevX.8.041031}
  {\bibfield  {journal} {\bibinfo  {journal} {Phys. Rev. X}\ }\textbf {\bibinfo
  {volume} {8}},\ \bibinfo {pages} {041031} (\bibinfo {year}
  {2018})}\BibitemShut {NoStop}%
\bibitem [{\citenamefont {Menssen}\ \emph {et~al.}(2020)\citenamefont
  {Menssen}, \citenamefont {Guan}, \citenamefont {Felce}, \citenamefont
  {Booth},\ and\ \citenamefont {Walmsley}}]{photonicmajorana3}%
  \BibitemOpen
  \bibfield  {author} {\bibinfo {author} {\bibfnamefont {A.~J.}\ \bibnamefont
  {Menssen}}, \bibinfo {author} {\bibfnamefont {J.}~\bibnamefont {Guan}},
  \bibinfo {author} {\bibfnamefont {D.}~\bibnamefont {Felce}}, \bibinfo
  {author} {\bibfnamefont {M.~J.}\ \bibnamefont {Booth}}, \ and\ \bibinfo
  {author} {\bibfnamefont {I.~A.}\ \bibnamefont {Walmsley}},\ }\href {\doibase
  10.1103/PhysRevLett.125.117401} {\bibfield  {journal} {\bibinfo  {journal}
  {Phys. Rev. Lett.}\ }\textbf {\bibinfo {volume} {125}},\ \bibinfo {pages}
  {117401} (\bibinfo {year} {2020})}\BibitemShut {NoStop}%
\bibitem [{\citenamefont {Goldman}\ \emph {et~al.}(2016)\citenamefont
  {Goldman}, \citenamefont {Budich},\ and\ \citenamefont
  {Zoller}}]{topocoldatomsreview}%
  \BibitemOpen
  \bibfield  {author} {\bibinfo {author} {\bibfnamefont {N.}~\bibnamefont
  {Goldman}}, \bibinfo {author} {\bibfnamefont {J.~C.}\ \bibnamefont {Budich}},
  \ and\ \bibinfo {author} {\bibfnamefont {P.}~\bibnamefont {Zoller}},\ }\href
  {\doibase 10.1038/nphys3803} {\bibfield  {journal} {\bibinfo  {journal}
  {Nature Physics}\ }\textbf {\bibinfo {volume} {12}},\ \bibinfo {pages} {639}
  (\bibinfo {year} {2016})}\BibitemShut {NoStop}%
\bibitem [{\citenamefont {Nascimb{\`{e}}ne}(2013)}]{Nascimb_ne_2013}%
  \BibitemOpen
  \bibfield  {author} {\bibinfo {author} {\bibfnamefont {S.}~\bibnamefont
  {Nascimb{\`{e}}ne}},\ }\href {\doibase 10.1088/0953-4075/46/13/134005}
  {\bibfield  {journal} {\bibinfo  {journal} {Journal of Physics B: Atomic,
  Molecular and Optical Physics}\ }\textbf {\bibinfo {volume} {46}},\ \bibinfo
  {pages} {134005} (\bibinfo {year} {2013})}\BibitemShut {NoStop}%
\bibitem [{\citenamefont {Jiang}\ \emph {et~al.}(2011)\citenamefont {Jiang},
  \citenamefont {Kitagawa}, \citenamefont {Alicea}, \citenamefont {Akhmerov},
  \citenamefont {Pekker}, \citenamefont {Refael}, \citenamefont {Cirac},
  \citenamefont {Demler}, \citenamefont {Lukin},\ and\ \citenamefont
  {Zoller}}]{majoranacoldatomwires}%
  \BibitemOpen
  \bibfield  {author} {\bibinfo {author} {\bibfnamefont {L.}~\bibnamefont
  {Jiang}}, \bibinfo {author} {\bibfnamefont {T.}~\bibnamefont {Kitagawa}},
  \bibinfo {author} {\bibfnamefont {J.}~\bibnamefont {Alicea}}, \bibinfo
  {author} {\bibfnamefont {A.~R.}\ \bibnamefont {Akhmerov}}, \bibinfo {author}
  {\bibfnamefont {D.}~\bibnamefont {Pekker}}, \bibinfo {author} {\bibfnamefont
  {G.}~\bibnamefont {Refael}}, \bibinfo {author} {\bibfnamefont {J.~I.}\
  \bibnamefont {Cirac}}, \bibinfo {author} {\bibfnamefont {E.}~\bibnamefont
  {Demler}}, \bibinfo {author} {\bibfnamefont {M.~D.}\ \bibnamefont {Lukin}}, \
  and\ \bibinfo {author} {\bibfnamefont {P.}~\bibnamefont {Zoller}},\ }\href
  {\doibase 10.1103/PhysRevLett.106.220402} {\bibfield  {journal} {\bibinfo
  {journal} {Phys. Rev. Lett.}\ }\textbf {\bibinfo {volume} {106}},\ \bibinfo
  {pages} {220402} (\bibinfo {year} {2011})}\BibitemShut {NoStop}%
\bibitem [{\citenamefont {Kraus}\ \emph {et~al.}(2013)\citenamefont {Kraus},
  \citenamefont {Dalmonte}, \citenamefont {Baranov}, \citenamefont
  {L\"auchli},\ and\ \citenamefont {Zoller}}]{PhysRevLett.111.173004}%
  \BibitemOpen
  \bibfield  {author} {\bibinfo {author} {\bibfnamefont {C.~V.}\ \bibnamefont
  {Kraus}}, \bibinfo {author} {\bibfnamefont {M.}~\bibnamefont {Dalmonte}},
  \bibinfo {author} {\bibfnamefont {M.~A.}\ \bibnamefont {Baranov}}, \bibinfo
  {author} {\bibfnamefont {A.~M.}\ \bibnamefont {L\"auchli}}, \ and\ \bibinfo
  {author} {\bibfnamefont {P.}~\bibnamefont {Zoller}},\ }\href {\doibase
  10.1103/PhysRevLett.111.173004} {\bibfield  {journal} {\bibinfo  {journal}
  {Phys. Rev. Lett.}\ }\textbf {\bibinfo {volume} {111}},\ \bibinfo {pages}
  {173004} (\bibinfo {year} {2013})}\BibitemShut {NoStop}%
\bibitem [{\citenamefont {Str\"ater}\ \emph {et~al.}(2016)\citenamefont
  {Str\"ater}, \citenamefont {Srivastava},\ and\ \citenamefont
  {Eckardt}}]{eckardtanyons}%
  \BibitemOpen
  \bibfield  {author} {\bibinfo {author} {\bibfnamefont {C.}~\bibnamefont
  {Str\"ater}}, \bibinfo {author} {\bibfnamefont {S.~C.~L.}\ \bibnamefont
  {Srivastava}}, \ and\ \bibinfo {author} {\bibfnamefont {A.}~\bibnamefont
  {Eckardt}},\ }\href {\doibase 10.1103/PhysRevLett.117.205303} {\bibfield
  {journal} {\bibinfo  {journal} {Phys. Rev. Lett.}\ }\textbf {\bibinfo
  {volume} {117}},\ \bibinfo {pages} {205303} (\bibinfo {year}
  {2016})}\BibitemShut {NoStop}%
\bibitem [{\citenamefont {Sachdev}\ \emph {et~al.}(2002)\citenamefont
  {Sachdev}, \citenamefont {Sengupta},\ and\ \citenamefont
  {Girvin}}]{sachdev-tiltedlatt1}%
  \BibitemOpen
  \bibfield  {author} {\bibinfo {author} {\bibfnamefont {S.}~\bibnamefont
  {Sachdev}}, \bibinfo {author} {\bibfnamefont {K.}~\bibnamefont {Sengupta}}, \
  and\ \bibinfo {author} {\bibfnamefont {S.~M.}\ \bibnamefont {Girvin}},\
  }\href {\doibase 10.1103/PhysRevB.66.075128} {\bibfield  {journal} {\bibinfo
  {journal} {Phys. Rev. B}\ }\textbf {\bibinfo {volume} {66}},\ \bibinfo
  {pages} {075128} (\bibinfo {year} {2002})}\BibitemShut {NoStop}%
\bibitem [{\citenamefont {Simon}\ \emph {et~al.}(2011)\citenamefont {Simon},
  \citenamefont {Bakr}, \citenamefont {Ma}, \citenamefont {Tai}, \citenamefont
  {Preiss},\ and\ \citenamefont {Greiner}}]{greiner-AFspinchains}%
  \BibitemOpen
  \bibfield  {author} {\bibinfo {author} {\bibfnamefont {J.}~\bibnamefont
  {Simon}}, \bibinfo {author} {\bibfnamefont {W.~S.}\ \bibnamefont {Bakr}},
  \bibinfo {author} {\bibfnamefont {R.}~\bibnamefont {Ma}}, \bibinfo {author}
  {\bibfnamefont {M.~E.}\ \bibnamefont {Tai}}, \bibinfo {author} {\bibfnamefont
  {P.~M.}\ \bibnamefont {Preiss}}, \ and\ \bibinfo {author} {\bibfnamefont
  {M.}~\bibnamefont {Greiner}},\ }\href {http://dx.doi.org/10.1038/nature09994}
  {\bibfield  {journal} {\bibinfo  {journal} {Nature}\ }\textbf {\bibinfo
  {volume} {472}},\ \bibinfo {pages} {307} (\bibinfo {year}
  {2011})}\BibitemShut {NoStop}%
\bibitem [{\citenamefont {Meinert}\ \emph {et~al.}(2013)\citenamefont
  {Meinert}, \citenamefont {Mark}, \citenamefont {Kirilov}, \citenamefont
  {Lauber}, \citenamefont {Weinmann}, \citenamefont {Daley},\ and\
  \citenamefont {N\"agerl}}]{nagerl-tiltedlatt}%
  \BibitemOpen
  \bibfield  {author} {\bibinfo {author} {\bibfnamefont {F.}~\bibnamefont
  {Meinert}}, \bibinfo {author} {\bibfnamefont {M.~J.}\ \bibnamefont {Mark}},
  \bibinfo {author} {\bibfnamefont {E.}~\bibnamefont {Kirilov}}, \bibinfo
  {author} {\bibfnamefont {K.}~\bibnamefont {Lauber}}, \bibinfo {author}
  {\bibfnamefont {P.}~\bibnamefont {Weinmann}}, \bibinfo {author}
  {\bibfnamefont {A.~J.}\ \bibnamefont {Daley}}, \ and\ \bibinfo {author}
  {\bibfnamefont {H.-C.}\ \bibnamefont {N\"agerl}},\ }\href {\doibase
  10.1103/PhysRevLett.111.053003} {\bibfield  {journal} {\bibinfo  {journal}
  {Phys. Rev. Lett.}\ }\textbf {\bibinfo {volume} {111}},\ \bibinfo {pages}
  {053003} (\bibinfo {year} {2013})}\BibitemShut {NoStop}%
\bibitem [{\citenamefont {Dimitrova}\ \emph {et~al.}(2020)\citenamefont
  {Dimitrova}, \citenamefont {Jepsen}, \citenamefont {Buyskikh}, \citenamefont
  {Venegas-Gomez}, \citenamefont {Amato-Grill}, \citenamefont {Daley},\ and\
  \citenamefont {Ketterle}}]{ketterle-tiltedlatt}%
  \BibitemOpen
  \bibfield  {author} {\bibinfo {author} {\bibfnamefont {I.}~\bibnamefont
  {Dimitrova}}, \bibinfo {author} {\bibfnamefont {N.}~\bibnamefont {Jepsen}},
  \bibinfo {author} {\bibfnamefont {A.}~\bibnamefont {Buyskikh}}, \bibinfo
  {author} {\bibfnamefont {A.}~\bibnamefont {Venegas-Gomez}}, \bibinfo {author}
  {\bibfnamefont {J.}~\bibnamefont {Amato-Grill}}, \bibinfo {author}
  {\bibfnamefont {A.}~\bibnamefont {Daley}}, \ and\ \bibinfo {author}
  {\bibfnamefont {W.}~\bibnamefont {Ketterle}},\ }\href {\doibase
  10.1103/PhysRevLett.124.043204} {\bibfield  {journal} {\bibinfo  {journal}
  {Phys. Rev. Lett.}\ }\textbf {\bibinfo {volume} {124}},\ \bibinfo {pages}
  {043204} (\bibinfo {year} {2020})}\BibitemShut {NoStop}%
\bibitem [{\citenamefont {Lieb}\ \emph {et~al.}(1961)\citenamefont {Lieb},
  \citenamefont {Schultz},\ and\ \citenamefont {Mattis}}]{LIEB1961407}%
  \BibitemOpen
  \bibfield  {author} {\bibinfo {author} {\bibfnamefont {E.}~\bibnamefont
  {Lieb}}, \bibinfo {author} {\bibfnamefont {T.}~\bibnamefont {Schultz}}, \
  and\ \bibinfo {author} {\bibfnamefont {D.}~\bibnamefont {Mattis}},\ }\href
  {\doibase https://doi.org/10.1016/0003-4916(61)90115-4} {\bibfield  {journal}
  {\bibinfo  {journal} {Annals of Physics}\ }\textbf {\bibinfo {volume} {16}},\
  \bibinfo {pages} {407 } (\bibinfo {year} {1961})}\BibitemShut {NoStop}%
\bibitem [{\citenamefont {DeGottardi}\ \emph {et~al.}(2013)\citenamefont
  {DeGottardi}, \citenamefont {Thakurathi}, \citenamefont {Vishveshwara},\ and\
  \citenamefont {Sen}}]{PhysRevB.88.165111}%
  \BibitemOpen
  \bibfield  {author} {\bibinfo {author} {\bibfnamefont {W.}~\bibnamefont
  {DeGottardi}}, \bibinfo {author} {\bibfnamefont {M.}~\bibnamefont
  {Thakurathi}}, \bibinfo {author} {\bibfnamefont {S.}~\bibnamefont
  {Vishveshwara}}, \ and\ \bibinfo {author} {\bibfnamefont {D.}~\bibnamefont
  {Sen}},\ }\href {\doibase 10.1103/PhysRevB.88.165111} {\bibfield  {journal}
  {\bibinfo  {journal} {Phys. Rev. B}\ }\textbf {\bibinfo {volume} {88}},\
  \bibinfo {pages} {165111} (\bibinfo {year} {2013})}\BibitemShut {NoStop}%
\bibitem [{\citenamefont {Kitaev}(2001)}]{kitaev1}%
  \BibitemOpen
  \bibfield  {author} {\bibinfo {author} {\bibfnamefont {A.~Y.}\ \bibnamefont
  {Kitaev}},\ }\href {http://stacks.iop.org/1063-7869/44/i=10S/a=S29}
  {\bibfield  {journal} {\bibinfo  {journal} {Physics-Uspekhi}\ }\textbf
  {\bibinfo {volume} {44}},\ \bibinfo {pages} {131} (\bibinfo {year}
  {2001})}\BibitemShut {NoStop}%
\bibitem [{\citenamefont {Zhang}\ \emph {et~al.}(2019)\citenamefont {Zhang},
  \citenamefont {Liu}, \citenamefont {Wimmer},\ and\ \citenamefont
  {Kouwenhoven}}]{Zhang:2019bl}%
  \BibitemOpen
  \bibfield  {author} {\bibinfo {author} {\bibfnamefont {H.}~\bibnamefont
  {Zhang}}, \bibinfo {author} {\bibfnamefont {D.~E.}\ \bibnamefont {Liu}},
  \bibinfo {author} {\bibfnamefont {M.}~\bibnamefont {Wimmer}}, \ and\ \bibinfo
  {author} {\bibfnamefont {L.~P.}\ \bibnamefont {Kouwenhoven}},\ }\href
  {\doibase 10.1038/s41467-019-13133-1} {\bibfield  {journal} {\bibinfo
  {journal} {Nature Communications}\ }\textbf {\bibinfo {volume} {10}},\
  \bibinfo {pages} {5128} (\bibinfo {year} {2019})}\BibitemShut {NoStop}%
\bibitem [{\citenamefont {Aasen}\ \emph {et~al.}(2016)\citenamefont {Aasen},
  \citenamefont {Hell}, \citenamefont {Mishmash}, \citenamefont {Higginbotham},
  \citenamefont {Danon}, \citenamefont {Leijnse}, \citenamefont {Jespersen},
  \citenamefont {Folk}, \citenamefont {Marcus}, \citenamefont {Flensberg},\
  and\ \citenamefont {Alicea}}]{PhysRevX.6.031016}%
  \BibitemOpen
  \bibfield  {author} {\bibinfo {author} {\bibfnamefont {D.}~\bibnamefont
  {Aasen}}, \bibinfo {author} {\bibfnamefont {M.}~\bibnamefont {Hell}},
  \bibinfo {author} {\bibfnamefont {R.~V.}\ \bibnamefont {Mishmash}}, \bibinfo
  {author} {\bibfnamefont {A.}~\bibnamefont {Higginbotham}}, \bibinfo {author}
  {\bibfnamefont {J.}~\bibnamefont {Danon}}, \bibinfo {author} {\bibfnamefont
  {M.}~\bibnamefont {Leijnse}}, \bibinfo {author} {\bibfnamefont {T.~S.}\
  \bibnamefont {Jespersen}}, \bibinfo {author} {\bibfnamefont {J.~A.}\
  \bibnamefont {Folk}}, \bibinfo {author} {\bibfnamefont {C.~M.}\ \bibnamefont
  {Marcus}}, \bibinfo {author} {\bibfnamefont {K.}~\bibnamefont {Flensberg}}, \
  and\ \bibinfo {author} {\bibfnamefont {J.}~\bibnamefont {Alicea}},\ }\href
  {\doibase 10.1103/PhysRevX.6.031016} {\bibfield  {journal} {\bibinfo
  {journal} {Phys. Rev. X}\ }\textbf {\bibinfo {volume} {6}},\ \bibinfo {pages}
  {031016} (\bibinfo {year} {2016})}\BibitemShut {NoStop}%
\bibitem [{\citenamefont {Sau}\ \emph {et~al.}(2020)\citenamefont {Sau},
  \citenamefont {Simon}, \citenamefont {Vishveshwara},\ and\ \citenamefont
  {Williams}}]{Sau:2020aj}%
  \BibitemOpen
  \bibfield  {author} {\bibinfo {author} {\bibfnamefont {J.}~\bibnamefont
  {Sau}}, \bibinfo {author} {\bibfnamefont {S.}~\bibnamefont {Simon}}, \bibinfo
  {author} {\bibfnamefont {S.}~\bibnamefont {Vishveshwara}}, \ and\ \bibinfo
  {author} {\bibfnamefont {J.~R.}\ \bibnamefont {Williams}},\ }\href {\doibase
  10.1038/s42254-020-00251-9} {\bibfield  {journal} {\bibinfo  {journal}
  {Nature Reviews Physics}\ } (\bibinfo {year} {2020}),\
  10.1038/s42254-020-00251-9}\BibitemShut {NoStop}%
\bibitem [{\citenamefont {Gemelke}\ \emph {et~al.}(2009)\citenamefont
  {Gemelke}, \citenamefont {Zhang}, \citenamefont {Hung},\ and\ \citenamefont
  {Chin}}]{chinmottplateaux}%
  \BibitemOpen
  \bibfield  {author} {\bibinfo {author} {\bibfnamefont {N.}~\bibnamefont
  {Gemelke}}, \bibinfo {author} {\bibfnamefont {X.}~\bibnamefont {Zhang}},
  \bibinfo {author} {\bibfnamefont {C.-L.}\ \bibnamefont {Hung}}, \ and\
  \bibinfo {author} {\bibfnamefont {C.}~\bibnamefont {Chin}},\ }\href@noop {}
  {\bibfield  {journal} {\bibinfo  {journal} {Nature}\ }\textbf {\bibinfo
  {volume} {460}},\ \bibinfo {pages} {995} (\bibinfo {year}
  {2009})}\BibitemShut {NoStop}%
\bibitem [{\citenamefont {Endres}\ \emph {et~al.}(2011)\citenamefont {Endres},
  \citenamefont {Cheneau}, \citenamefont {Fukuhara}, \citenamefont
  {Weitenberg}, \citenamefont {Schau{\ss}}, \citenamefont {Gross},
  \citenamefont {Mazza}, \citenamefont {Ba{\~n}uls}, \citenamefont {Pollet},
  \citenamefont {Bloch},\ and\ \citenamefont {Kuhr}}]{bloch-stringorder}%
  \BibitemOpen
  \bibfield  {author} {\bibinfo {author} {\bibfnamefont {M.}~\bibnamefont
  {Endres}}, \bibinfo {author} {\bibfnamefont {M.}~\bibnamefont {Cheneau}},
  \bibinfo {author} {\bibfnamefont {T.}~\bibnamefont {Fukuhara}}, \bibinfo
  {author} {\bibfnamefont {C.}~\bibnamefont {Weitenberg}}, \bibinfo {author}
  {\bibfnamefont {P.}~\bibnamefont {Schau{\ss}}}, \bibinfo {author}
  {\bibfnamefont {C.}~\bibnamefont {Gross}}, \bibinfo {author} {\bibfnamefont
  {L.}~\bibnamefont {Mazza}}, \bibinfo {author} {\bibfnamefont {M.~C.}\
  \bibnamefont {Ba{\~n}uls}}, \bibinfo {author} {\bibfnamefont
  {L.}~\bibnamefont {Pollet}}, \bibinfo {author} {\bibfnamefont
  {I.}~\bibnamefont {Bloch}}, \ and\ \bibinfo {author} {\bibfnamefont
  {S.}~\bibnamefont {Kuhr}},\ }\href@noop {} {\bibfield  {journal} {\bibinfo
  {journal} {Science}\ }\textbf {\bibinfo {volume} {334}},\ \bibinfo {pages}
  {200} (\bibinfo {year} {2011})}\BibitemShut {NoStop}%
\bibitem [{\citenamefont {Campbell}\ \emph {et~al.}(2006)\citenamefont
  {Campbell}, \citenamefont {Mun}, \citenamefont {Boyd}, \citenamefont
  {Medley}, \citenamefont {Leanhardt}, \citenamefont {Marcassa}, \citenamefont
  {Pritchard},\ and\ \citenamefont {Ketterle}}]{clockshiftmott}%
  \BibitemOpen
  \bibfield  {author} {\bibinfo {author} {\bibfnamefont {G.~K.}\ \bibnamefont
  {Campbell}}, \bibinfo {author} {\bibfnamefont {J.}~\bibnamefont {Mun}},
  \bibinfo {author} {\bibfnamefont {M.}~\bibnamefont {Boyd}}, \bibinfo {author}
  {\bibfnamefont {P.}~\bibnamefont {Medley}}, \bibinfo {author} {\bibfnamefont
  {A.~E.}\ \bibnamefont {Leanhardt}}, \bibinfo {author} {\bibfnamefont {L.~G.}\
  \bibnamefont {Marcassa}}, \bibinfo {author} {\bibfnamefont {D.~E.}\
  \bibnamefont {Pritchard}}, \ and\ \bibinfo {author} {\bibfnamefont
  {W.}~\bibnamefont {Ketterle}},\ }\href
  {http://www.sciencemag.org/cgi/content/abstract/313/5787/649} {\bibfield
  {journal} {\bibinfo  {journal} {Science}\ }\textbf {\bibinfo {volume}
  {313}},\ \bibinfo {pages} {649} (\bibinfo {year} {2006})}\BibitemShut
  {NoStop}%
\bibitem [{\citenamefont {St\"oferle}\ \emph {et~al.}(2004)\citenamefont
  {St\"oferle}, \citenamefont {Moritz}, \citenamefont {Schori}, \citenamefont
  {K\"ohl},\ and\ \citenamefont
  {Esslinger}}]{esslinger_SMFI_latticemodulation}%
  \BibitemOpen
  \bibfield  {author} {\bibinfo {author} {\bibfnamefont {T.}~\bibnamefont
  {St\"oferle}}, \bibinfo {author} {\bibfnamefont {H.}~\bibnamefont {Moritz}},
  \bibinfo {author} {\bibfnamefont {C.}~\bibnamefont {Schori}}, \bibinfo
  {author} {\bibfnamefont {M.}~\bibnamefont {K\"ohl}}, \ and\ \bibinfo {author}
  {\bibfnamefont {T.}~\bibnamefont {Esslinger}},\ }\href@noop {} {\bibfield
  {journal} {\bibinfo  {journal} {Phys. Rev. Lett.}\ }\textbf {\bibinfo
  {volume} {92}},\ \bibinfo {pages} {130403} (\bibinfo {year}
  {2004})}\BibitemShut {NoStop}%
\bibitem [{\citenamefont {Amato-Grill}\ \emph {et~al.}(2019)\citenamefont
  {Amato-Grill}, \citenamefont {Jepsen}, \citenamefont {Dimitrova},
  \citenamefont {Lunden},\ and\ \citenamefont
  {Ketterle}}]{KetterleMottspectroscopy}%
  \BibitemOpen
  \bibfield  {author} {\bibinfo {author} {\bibfnamefont {J.}~\bibnamefont
  {Amato-Grill}}, \bibinfo {author} {\bibfnamefont {N.}~\bibnamefont {Jepsen}},
  \bibinfo {author} {\bibfnamefont {I.}~\bibnamefont {Dimitrova}}, \bibinfo
  {author} {\bibfnamefont {W.}~\bibnamefont {Lunden}}, \ and\ \bibinfo {author}
  {\bibfnamefont {W.}~\bibnamefont {Ketterle}},\ }\href {\doibase
  10.1103/PhysRevA.99.033612} {\bibfield  {journal} {\bibinfo  {journal} {Phys.
  Rev. A}\ }\textbf {\bibinfo {volume} {99}},\ \bibinfo {pages} {033612}
  (\bibinfo {year} {2019})}\BibitemShut {NoStop}%
\bibitem [{\citenamefont {Cl\'ement}\ \emph {et~al.}(2009)\citenamefont
  {Cl\'ement}, \citenamefont {Fabbri}, \citenamefont {Fallani}, \citenamefont
  {Fort},\ and\ \citenamefont {Inguscio}}]{inguscio-bragg}%
  \BibitemOpen
  \bibfield  {author} {\bibinfo {author} {\bibfnamefont {D.}~\bibnamefont
  {Cl\'ement}}, \bibinfo {author} {\bibfnamefont {N.}~\bibnamefont {Fabbri}},
  \bibinfo {author} {\bibfnamefont {L.}~\bibnamefont {Fallani}}, \bibinfo
  {author} {\bibfnamefont {C.}~\bibnamefont {Fort}}, \ and\ \bibinfo {author}
  {\bibfnamefont {M.}~\bibnamefont {Inguscio}},\ }\href {\doibase
  10.1103/PhysRevLett.102.155301} {\bibfield  {journal} {\bibinfo  {journal}
  {Phys. Rev. Lett.}\ }\textbf {\bibinfo {volume} {102}},\ \bibinfo {pages}
  {155301} (\bibinfo {year} {2009})}\BibitemShut {NoStop}%
\bibitem [{\citenamefont {Ernst}\ \emph {et~al.}(2010)\citenamefont {Ernst},
  \citenamefont {G{\"o}tze}, \citenamefont {Krauser}, \citenamefont {Pyka},
  \citenamefont {L{\"u}hmann}, \citenamefont {Pfannkuche},\ and\ \citenamefont
  {Sengstock}}]{sengstock-bragg}%
  \BibitemOpen
  \bibfield  {author} {\bibinfo {author} {\bibfnamefont {P.~T.}\ \bibnamefont
  {Ernst}}, \bibinfo {author} {\bibfnamefont {S.}~\bibnamefont {G{\"o}tze}},
  \bibinfo {author} {\bibfnamefont {J.~S.}\ \bibnamefont {Krauser}}, \bibinfo
  {author} {\bibfnamefont {K.}~\bibnamefont {Pyka}}, \bibinfo {author}
  {\bibfnamefont {D.-S.}\ \bibnamefont {L{\"u}hmann}}, \bibinfo {author}
  {\bibfnamefont {D.}~\bibnamefont {Pfannkuche}}, \ and\ \bibinfo {author}
  {\bibfnamefont {K.}~\bibnamefont {Sengstock}},\ }\href {\doibase
  10.1038/nphys1476} {\bibfield  {journal} {\bibinfo  {journal} {Nature
  Physics}\ }\textbf {\bibinfo {volume} {6}},\ \bibinfo {pages} {56} (\bibinfo
  {year} {2010})}\BibitemShut {NoStop}%
\bibitem [{\citenamefont {Geiger}\ \emph {et~al.}(2018)\citenamefont {Geiger},
  \citenamefont {Fujiwara}, \citenamefont {Singh}, \citenamefont {Senaratne},
  \citenamefont {Rajagopal}, \citenamefont {Lipatov}, \citenamefont
  {Shimasaki}, \citenamefont {Driben}, \citenamefont {Konotop}, \citenamefont
  {Meier},\ and\ \citenamefont {Weld}}]{RSBO-PRL}%
  \BibitemOpen
  \bibfield  {author} {\bibinfo {author} {\bibfnamefont {Z.~A.}\ \bibnamefont
  {Geiger}}, \bibinfo {author} {\bibfnamefont {K.~M.}\ \bibnamefont
  {Fujiwara}}, \bibinfo {author} {\bibfnamefont {K.}~\bibnamefont {Singh}},
  \bibinfo {author} {\bibfnamefont {R.}~\bibnamefont {Senaratne}}, \bibinfo
  {author} {\bibfnamefont {S.~V.}\ \bibnamefont {Rajagopal}}, \bibinfo {author}
  {\bibfnamefont {M.}~\bibnamefont {Lipatov}}, \bibinfo {author} {\bibfnamefont
  {T.}~\bibnamefont {Shimasaki}}, \bibinfo {author} {\bibfnamefont
  {R.}~\bibnamefont {Driben}}, \bibinfo {author} {\bibfnamefont {V.~V.}\
  \bibnamefont {Konotop}}, \bibinfo {author} {\bibfnamefont {T.}~\bibnamefont
  {Meier}}, \ and\ \bibinfo {author} {\bibfnamefont {D.~M.}\ \bibnamefont
  {Weld}},\ }\href {\doibase 10.1103/PhysRevLett.120.213201} {\bibfield
  {journal} {\bibinfo  {journal} {Phys. Rev. Lett.}\ }\textbf {\bibinfo
  {volume} {120}},\ \bibinfo {pages} {213201} (\bibinfo {year}
  {2018})}\BibitemShut {NoStop}%
\bibitem [{\citenamefont {Fujiwara}\ \emph {et~al.}(2019)\citenamefont
  {Fujiwara}, \citenamefont {Singh}, \citenamefont {Geiger}, \citenamefont
  {Senaratne}, \citenamefont {Rajagopal}, \citenamefont {Lipatov},\ and\
  \citenamefont {Weld}}]{warpdrivePRL}%
  \BibitemOpen
  \bibfield  {author} {\bibinfo {author} {\bibfnamefont {C.~J.}\ \bibnamefont
  {Fujiwara}}, \bibinfo {author} {\bibfnamefont {K.}~\bibnamefont {Singh}},
  \bibinfo {author} {\bibfnamefont {Z.~A.}\ \bibnamefont {Geiger}}, \bibinfo
  {author} {\bibfnamefont {R.}~\bibnamefont {Senaratne}}, \bibinfo {author}
  {\bibfnamefont {S.~V.}\ \bibnamefont {Rajagopal}}, \bibinfo {author}
  {\bibfnamefont {M.}~\bibnamefont {Lipatov}}, \ and\ \bibinfo {author}
  {\bibfnamefont {D.~M.}\ \bibnamefont {Weld}},\ }\href {\doibase
  10.1103/PhysRevLett.122.010402} {\bibfield  {journal} {\bibinfo  {journal}
  {Phys. Rev. Lett.}\ }\textbf {\bibinfo {volume} {122}},\ \bibinfo {pages}
  {010402} (\bibinfo {year} {2019})}\BibitemShut {NoStop}%
\bibitem [{\citenamefont {Rajagopal}\ \emph {et~al.}(2019)\citenamefont
  {Rajagopal}, \citenamefont {Shimasaki}, \citenamefont {Dotti}, \citenamefont
  {Raciunas}, \citenamefont {Senaratne}, \citenamefont {Anisimovas},
  \citenamefont {Eckardt},\ and\ \citenamefont {Weld}}]{phasonPRL}%
  \BibitemOpen
  \bibfield  {author} {\bibinfo {author} {\bibfnamefont {S.~V.}\ \bibnamefont
  {Rajagopal}}, \bibinfo {author} {\bibfnamefont {T.}~\bibnamefont
  {Shimasaki}}, \bibinfo {author} {\bibfnamefont {P.}~\bibnamefont {Dotti}},
  \bibinfo {author} {\bibfnamefont {M.}~\bibnamefont {Raciunas}}, \bibinfo
  {author} {\bibfnamefont {R.}~\bibnamefont {Senaratne}}, \bibinfo {author}
  {\bibfnamefont {E.}~\bibnamefont {Anisimovas}}, \bibinfo {author}
  {\bibfnamefont {A.}~\bibnamefont {Eckardt}}, \ and\ \bibinfo {author}
  {\bibfnamefont {D.~M.}\ \bibnamefont {Weld}},\ }\href@noop {} {\bibfield
  {journal} {\bibinfo  {journal} {Phys. Rev. Lett.}\ }\textbf {\bibinfo
  {volume} {123}} (\bibinfo {year} {2019})}\BibitemShut {NoStop}%
\bibitem [{\citenamefont {Gorantla}\ \emph {et~al.}(2020)\citenamefont
  {Gorantla}, \citenamefont {Lam}, \citenamefont {Seiberg},\ and\ \citenamefont
  {Shao}}]{seibergarxiv}%
  \BibitemOpen
  \bibfield  {author} {\bibinfo {author} {\bibfnamefont {P.}~\bibnamefont
  {Gorantla}}, \bibinfo {author} {\bibfnamefont {H.~T.}\ \bibnamefont {Lam}},
  \bibinfo {author} {\bibfnamefont {N.}~\bibnamefont {Seiberg}}, \ and\
  \bibinfo {author} {\bibfnamefont {S.-H.}\ \bibnamefont {Shao}},\ }\href@noop
  {} {\bibfield  {journal} {\bibinfo  {journal} {arXiv:2010.16414}\ } (\bibinfo
  {year} {2020})}\BibitemShut {NoStop}%
\end{thebibliography}

%

\bibliographystyle{apsrev4-1}

\clearpage

\end{document}